\documentstyle[amssymb,aps,preprint]{revtex}

\begin{document}
\title{Sum Rule for the Optical Absorption of an Interacting Many-Polaron Gas}
\author{J.\ Tempere and J. T. Devreese}
\address{Departement Natuurkunde, Universiteit Antwerpen - UIA, Universiteitsplein 1,
B2610 Antwerpen, Belgium.}
\date{1/12/2000}
\maketitle

\begin{abstract}
A sum rule for the first frequency moment of the optical absorption of a
many-polaron system is derived, taking into account many-body effects in the
system of constituent charge carriers of the many-polaron system. In our
expression for the sum rule, the electron-phonon coupling and the many-body
effects in the electron (or hole) system formally decouple, so that the
many-body effects can be treated to the desired level of approximation by
the choice of the dynamical structure factor of the electron (hole) gas. We
calculate correction factors to take into account both low and high
experimental cutoff frequencies. 
\end{abstract}


\section{Introduction}

Sum rules have proven to be a powerful tool in the analysis of optical
spectra \cite{BrauwersPRB12}. They provide expressions relating a frequency
moment of the optical absorption spectrum to characteristic properties of a
material such as its plasma frequency. Sum rules provide useful checks (on
the optical properties) both for theory and for experiment. 

The goal of this paper is to derive a sum rule for the normalized first
frequency moment of the optical absorption of a gas of continuum (Fr\"{o}%
hlich) polarons, including many-body effects between the charge carriers in
this many-polaron gas. The normalized first frequency moment of the optical
absorption is obtained from the optical conductivity $%
\mathop{\rm Re}%
[\sigma (\omega )]$ by 
\begin{equation}
\left\langle \omega \right\rangle =%
{\displaystyle{%
\displaystyle\int \limits_{0}^{\infty }\omega %
\mathop{\rm Re}[\sigma (\omega )] \over %
\displaystyle\int \limits_{0}^{\infty }%
\mathop{\rm Re}[\sigma (\omega )]d\omega }}%
.  \label{<w>}
\end{equation}
This quantity is experimentally accessible: for example Calvani and
co-workers\ \cite{LupiPRL83}\ have determined $\left\langle \omega
\right\rangle $ as a function of doping (density) for the optical absorption
bands{\bf \ }attributed to polarons in a family of high-temperature cuprate
superconductors \cite{LupiPRL83}. As such, the sum rule (\ref{<w>}) for $%
\left\langle \omega \right\rangle $ applied to the polaron gas including
many-body effects (such as interaction, screening, occupational effects of
Fermi statistics,...) will prove a useful tool for the analysis of infrared
spectra of such high-T$_{c}$ materials. The study of infrared spectra in the
framework of polaron theory can complement other indications of the presence
of Fr\"{o}hlich polarons and bipolarons \cite{AlexandrovPRL77} in high-T$_{c}
$ materials, thus providing a more solid ground for hypotheses involving
polarons and bipolarons in the mechanism of high-temperature
superconductivity \cite{MottPHY205}.

In the derivation presented here (in Section II), the many-body effects in
the system of charge carriers are completely contained in the dynamical
structure factor of the charge carrier gas, and they are formally factored
out from the charge carrier - phonon interactions, which allows for a great
deal of flexibility in treating the many-polaron gas.

We demonstrate how to circumvent two particular difficulties which
complicate the practical use of many-polaron sum rules in a comparison with
experiment. The first difficulty is related to the presence of a delta
function contribution at zero frequency and $T=0$\ in the theoretical
optical absorption spectrum of the polaron gas, at temperature zero \cite
{DevreesePRB15}. Such a sharp zero-frequency feature can be overlooked by
experiment, and we derive a correction factor --which turns out to be
related to the density dependent effective polaron mass-- to compensate for
the inability to detect the delta function contribution. This correction
factor is the subject of Section III. A second difficulty is that the
theoretical many-polaron optical absorption spectrum contains a slowly
decaying high-frequency tail which carries a substantial fraction of the
spectral weight of the polaron optical absorption. In Section IV we discuss
the influence of a high-frequency cutoff on the first frequency moment and
propose a formula useful to extrapolate experimental data of the polaron
optical absorption band at higher frequencies.

\section{First frequency moment of the optical absorption of a many-polaron
gas}

\subsection{Exact expression}

The optical absorption coefficient is proportional to the real part of the
optical conductivity, $%
\mathop{\rm Re}%
[\sigma (\omega )]$. To evaluate expression (\ref{<w>}), both the zeroth and
the first frequency moment of the real part of the optical conductivity have
to be determined. The zeroth moment is easily related to the plasma
frequency through the f-sum rule \cite{pcmartin} : 
\begin{equation}
\displaystyle\int %
\limits_{0}^{\infty }%
\mathop{\rm Re}%
[\sigma (\omega )]d\omega =\frac{\pi Ne^{2}}{2m_{b}}.  \label{zeroeth}
\end{equation}
In this expression, $N$ is the number of polarons per unit volume, $m_{b}$
is the band mass of the constituent electron (or hole) of the polaron, and $e
$ is the electron charge. To obtain the first frequency moment of $%
\mathop{\rm Re}%
[\sigma (\omega )]$, the optical conductivity is expressed as a
current-current correlation function according to the Kubo formalism of
linear response theory (see e.g. \cite{mahan}). Integrating twice by parts,
the current-current correlation function can be transformed into a
force-force correlation function \cite{Greenbook2}. This leads to: 
\begin{equation}
\mathop{\rm Re}%
[\sigma (\omega )]=\frac{1}{\text{V}\hbar \omega ^{3}}\frac{e^{2}}{m_{b}^{2}}%
\mathop{\rm Re}%
\left\{ \int_{0}^{\infty }e^{i\omega t}\left\langle \left[ F_{x}(t),F_{x}(0)%
\right] \right\rangle dt\right\} .  \label{resig}
\end{equation}
where V is the volume (chosen equal to unity in what follows) and $F_{x}$ is
the $x$-component of the force operator ${\bf F}=(i/\hbar )\left[
H,\sum_{j=1}^{N}{\bf p}_{j}\right] $. To formulate the sum rule, the
force-force correlation function is rewritten in spectral representation.
For this purpose, we introduce a set $\{\Psi _{n}\}$ of basis states with
corresponding energies $\{E_{n}\},$ of which $\Psi _{0}$ is the ground state
with ground state energy $E_{0}$. The spectral representation then becomes: 
\begin{eqnarray}
\int_{0}^{\infty }e^{i\omega t}\left\langle \left[ F_{x}(t),F_{x}(0)\right]
\right\rangle dt &=&\sum_{n}\int_{0}^{\infty }e^{i\omega t-\delta t}\left[ 
\begin{array}{c}
\left\langle \Psi _{0}\left| e^{-iHt/\hbar }F_{x}e^{iHt/\hbar }\right| \Psi
_{n}\right\rangle \left\langle \Psi _{n}\left| F_{x}\right| \Psi
_{0}\right\rangle  \\ 
-\left\langle \Psi _{0}\left| F_{x}\right| \Psi _{n}\right\rangle
\left\langle \Psi _{n}\left| e^{-iHt/\hbar }F_{x}e^{iHt/\hbar }\right| \Psi
_{0}\right\rangle 
\end{array}
\right] dt \\
&=&\sum_{n}i\left| \left\langle \Psi _{0}\left| F_{x}\right| \Psi
_{n}\right\rangle \right| ^{2}\left( 
\begin{array}{c}
{\displaystyle{1 \over \omega +(E_{n}-E_{0})/\hbar +i\delta }}%
\\ 
-%
{\displaystyle{1 \over \omega -(E_{n}-E_{0})/\hbar +i\delta }}%
\end{array}
\right) .
\end{eqnarray}
Denoting $(E_{n}-E_{0})/\hbar $ as $\omega _{n0}$, one arrives at the
following expression for the first moment of the optical conductivity: 
\begin{eqnarray}
\displaystyle\int %
\limits_{0}^{\infty }\omega 
\mathop{\rm Re}%
[\sigma (\omega )] &=&%
\displaystyle\int %
\limits_{0}^{\infty }d\omega \frac{e^{2}}{m_{b}^{2}\hbar \omega ^{2}}%
\mathop{\rm Re}%
\left\{ \sum_{n}i\left| \left\langle \Psi _{0}\left| F_{x}\right| \Psi
_{n}\right\rangle \right| ^{2}\left( 
\begin{array}{c}
{\displaystyle{1 \over \omega +(E_{n}-E_{0})/\hbar +i\delta }}%
\\ 
-%
{\displaystyle{1 \over \omega -(E_{n}-E_{0})/\hbar +i\delta }}%
\end{array}
\right) \right\}  \\
&=&\frac{\pi e^{2}}{m_{b}^{2}\hbar }\sum_{n}%
{\displaystyle{\left| \left\langle \Psi _{0}\left| F_{x}\right| \Psi _{n}\right\rangle \right| ^{2} \over \omega _{n0}^{2}}}%
.  \label{first}
\end{eqnarray}
The result for the normalized first frequency moment of the optical
conductivity is found from the expressions (\ref{first}) and (\ref{zeroeth}%
): 
\begin{equation}
\left\langle \omega \right\rangle =%
{\displaystyle{%
\displaystyle\int \limits_{0}^{\infty }\omega %
\mathop{\rm Re}[\sigma (\omega )] \over %
\displaystyle\int \limits_{0}^{\infty }%
\mathop{\rm Re}[\sigma (\omega )]d\omega }}%
=\frac{2}{N\hbar m_{b}}\sum_{n}\left| 
{\displaystyle{\left\langle \Psi _{0}\left| F_{x}\right| \Psi _{n}\right\rangle  \over \omega _{n0}}}%
\right| ^{2}.  \label{firmom1}
\end{equation}
To obtain a useful expression, the resummation over the unspecified set of
basis functions $\{\Psi _{n}\}$ should still be performed. This is done by
using the integral representation 
\begin{equation}
{\displaystyle{1 \over \omega _{n0}}}%
=\lim\limits_{\delta \rightarrow +0}\left( i%
\displaystyle\int %
\limits_{0}^{\infty }e^{i\omega _{n0}t-\delta t}dt\right) 
\end{equation}
such that 
\begin{eqnarray}
\left\langle \omega \right\rangle  &=&\frac{2}{N\hbar m_{b}}%
\lim\limits_{\delta \rightarrow +0}%
\displaystyle\int %
\limits_{0}^{\infty }dt%
\displaystyle\int %
\limits_{0}^{\infty }ds\text{ }\sum_{n}\left\langle \Psi _{0}\left|
F_{x}(t)\right| \Psi _{n}\right\rangle \left\langle \Psi _{n}\left|
F_{x}(s)\right| \Psi _{0}\right\rangle e^{-\delta (t+s)} \\
&=&\frac{2}{N\hbar m_{b}}\lim\limits_{\delta \rightarrow +0}2%
\mathop{\rm Re}%
\left[ 
\displaystyle\int %
\limits_{0}^{\infty }dt%
\displaystyle\int %
\limits_{0}^{t}ds\left\langle \Psi _{0}\left| F_{x}(t)F_{x}(s)\right| \Psi
_{0}\right\rangle e^{-\delta (t+s)}\right] ,
\end{eqnarray}
where the Hermiticity of the force operator was used. If the response is
time translational invariant (meaning that the response is only dependent on
the time lapse since perturbation), the final result can be written as: 
\begin{equation}
\left\langle \omega \right\rangle =\frac{2}{N\hbar m_{b}}\lim\limits_{\delta
\rightarrow +0}\left[ 
{\displaystyle{1 \over \delta }}%
\displaystyle\int %
\limits_{0}^{\infty }\left\langle \Psi _{0}\left| F_{x}(t)F_{x}(0)\right|
\Psi _{0}\right\rangle e^{-\delta t}dt\right] .  \label{firmom2}
\end{equation}
Expression (\ref{firmom2}) is an exact expression, within the assumptions of
linear response theory, for the sum rule for the normalized first frequency
moment of the optical absorption of the many-polaron gas.

\subsection{The interacting Fr\"{o}hlich many-polaron gas}

Throughout this paper, we consider polarons in the continuum or Fr\"{o}hlich
approximation \cite{frohlich}. The system of $N$ interacting Fr\"{o}hlich
polarons is then described by the Hamiltonian 
\begin{equation}
H=\sum_{j=1}^{N}%
{\displaystyle{p_{j}^{2} \over 2m_{b}}}%
+\sum_{{\bf k}}\hbar \omega _{\text{LO}}b_{{\bf k}}^{+}b_{{\bf k}}+\sum_{%
{\bf k}}\left( V_{{\bf k}}b_{{\bf k}}\rho _{{\bf k}}+V_{{\bf k}}^{\ast }b_{%
{\bf k}}^{+}\rho _{{\bf k}}^{+}\right) +\sum_{j=1}^{N}\sum_{l\neq j=1}^{N}%
{\displaystyle{e^{2}/\varepsilon _{\infty } \over |{\bf r}_{j}-{\bf r}_{l}|}}%
,
\end{equation}
where ${\bf r}_{j},{\bf p}_{j}$ are the position and momentum operators of
charge carrier $j$ with band mass $m_{b}$ and change $\pm e$, and $a_{{\bf k}%
}^{+}$,$a_{{\bf k}}$ are the creation and annihilation operators of a
longitudinal optical (LO) phonon with frequency $\omega _{\text{LO}}$, $\rho
_{{\bf k}}=\sum_{j=1}^{N}\exp \{i{\bf k}.{\bf r}_{j}\}$ is the density
operator of the charge carriers, $V_{{\bf k}}$ is the interaction amplitude
matrix element characterizing the interaction between the charge carriers
and the LO phonons, and $\varepsilon _{\infty }$ is the permittivity of the
medium. The total force operator for the many-polaron gas, with both charge
carrier - phonon interaction and Coulomb interaction between the charge
carriers taken into account, takes the form 
\[
{\bf F}=%
{\displaystyle{i \over \hbar }}%
\left[ H,\sum_{j=1}^{N}{\bf p}_{j}\right] =-i\sum_{{\bf k}}{\bf k}(V_{{\bf k}%
}b_{{\bf k}}\rho _{{\bf k}}-V_{{\bf k}}^{\ast }b_{{\bf k}}^{+}\rho _{{\bf k}%
}^{+}),
\]
so that in the force-force correlation $F_{x}(t)F_{x}(0)$, a factor $|V_{%
{\bf k}}|^{2}\propto \alpha $\ is present and formally factors out of the
expression for $\left\langle \omega \right\rangle $\ and $%
\mathop{\rm Re}%
[\sigma (\omega )]$\ in equations (\ref{firmom2}) and (\ref{resig}),
respectively. Herein lies an advantage of working with the force-force
correlation as compared to the current-current correlation, especially at
weak coupling. Thus, to calculate $\left\langle \omega \right\rangle $ (\ref
{firmom2}) to lowest order in the charge carrier - phonon interaction
strength $\alpha $ it is sufficient to evaluate the ($\alpha $-independent)
expectation value of the force-force correlation in a product wave function $%
\left| \phi \right\rangle \left| \varphi _{\text{el}}\right\rangle $ where $%
\left| \varphi _{\text{el}}\right\rangle $ represents the ground-state wave
function for charge carriers and $\left| \phi \right\rangle $ is the phonon
vacuum: 
\begin{eqnarray}
\left\langle \omega \right\rangle  &=&\frac{2}{N\hbar m_{b}}%
\lim\limits_{\delta \rightarrow +0}\left\{ 
{\displaystyle{1 \over \delta }}%
\displaystyle\int %
\limits_{0}^{\infty }dte^{-\delta t}\sum_{{\bf k,k}^{\prime
}}k_{x}.k_{x}^{\prime }\right.   \nonumber \\
&&\left. \times \left\langle \varphi _{\text{el}}\left| \left\langle \phi
\left| 
\begin{array}{c}
\left[ V_{{\bf k}}b_{{\bf k}}(t)\rho _{{\bf k}}(t)-V_{{\bf k}^{\prime
}}^{\ast }b_{{\bf k}^{\prime }}^{+}(t)\rho _{{\bf k}^{\prime }}^{+}(t)\right]
\\ 
\times \left[ V_{{\bf k}^{\prime }}b_{{\bf k}^{\prime }}(0)\rho _{{\bf k}%
^{\prime }}(0)-V_{{\bf k}^{\prime }}^{\ast }b_{{\bf k}^{\prime }}^{+}(0)\rho
_{{\bf k}^{\prime }}^{+}(0)\right] 
\end{array}
\right| \phi \right\rangle \right| \varphi _{\text{el}}\right\rangle
\right\} .  \label{<w>bis}
\end{eqnarray}
The expectation value with respect to the phonon vacuum can be calculated to
order $\alpha $ and results in 
\begin{equation}
\left\langle \omega \right\rangle =-\frac{2}{N\hbar m_{b}}\sum_{{\bf k}%
}k_{x}^{2}\left| V_{{\bf k}}\right| ^{2}\lim\limits_{\delta \rightarrow
+0}\left\{ 
{\displaystyle{1 \over \delta }}%
\displaystyle\int %
\limits_{0}^{\infty }dt\text{ }e^{i\omega _{\text{LO}}t}e^{-\delta
t}\left\langle \varphi _{\text{el}}\left| \rho _{{\bf k}}(t)\rho _{{\bf k}%
}^{+}(0)\right| \varphi _{\text{el}}\right\rangle \right\} .
\end{equation}
The remaining expectation value is the density-density Green's function $G(%
{\bf k},t)=-i\Theta (t)\left\langle \varphi _{\text{el}}\left| \rho _{{\bf k}%
}(t)\rho _{{\bf k}}^{+}(0)\right| \varphi _{\text{el}}\right\rangle /N$ so
that 
\begin{equation}
\left\langle \omega \right\rangle =-\frac{2}{\hbar m_{b}}\sum_{{\bf k}%
}k_{x}^{2}\left| V_{{\bf k}}\right| ^{2}\lim\limits_{\delta \rightarrow
+0}\left\{ 
{\displaystyle{1 \over \delta }}%
G({\bf k},-\omega _{\text{LO}}+i\delta )\right\} .
\end{equation}
Introducing the dynamical structure factor $S(k,\omega )$ as the spectral
density function of the retarded density-density Green's function $G_{\text{R%
}}$ of the electron (or hole) gas, we arrive at: 
\begin{equation}
\left\langle \omega \right\rangle =-\frac{2}{\hbar m_{b}}\sum_{{\bf k}%
}k_{x}^{2}\left| V_{{\bf k}}\right| ^{2}%
\displaystyle\int %
\limits_{0}^{\infty }d\omega \text{ }%
{\displaystyle{S({\bf k},\omega -\omega _{\text{LO}}) \over \omega ^{2}}}%
.
\end{equation}
The modulus square of the Fr\"{o}hlich interaction amplitude is given by 
\begin{equation}
|V_{{\bf k}}|^{2}=\left\{ 
\begin{array}{l}
{\displaystyle{(\hbar \omega _{\text{LO}})^{2} \over k^{2}}}%
{\displaystyle{4\pi \alpha  \over \text{V}}}%
\sqrt{%
{\displaystyle{\hbar  \over 2m_{b}\omega _{\text{LO}}}}%
}\text{ in 3D} \\ 
{\displaystyle{(\hbar \omega _{\text{LO}})^{2} \over k}}%
{\displaystyle{2\pi \alpha  \over \text{A}}}%
\sqrt{%
{\displaystyle{\hbar  \over 2m_{b}\omega _{\text{LO}}}}%
}\,\text{in 2D,}
\end{array}
\right. 
\end{equation}
where V (A) is the volume (surface). This leads to 
\begin{equation}
\left\langle \omega \right\rangle _{\text{3D}}=-\alpha \omega _{\text{LO}}%
{\displaystyle{\hbar \omega _{\text{LO}} \over m_{b}}}%
\sqrt{%
{\displaystyle{\hbar  \over m_{b}\omega _{\text{LO}}}}%
}%
{\displaystyle{2\sqrt{2} \over 3\pi }}%
\displaystyle\int %
\limits_{0}^{\infty }dk%
\displaystyle\int %
\limits_{0}^{\infty }d\omega \text{ }k^{2}%
{\displaystyle{S_{\text{3D}}({\bf k},\omega -\omega _{\text{LO}}) \over \omega ^{2}}}%
,  \eqnum{19a}  \label{w3d}
\end{equation}
and 
\begin{equation}
\left\langle \omega \right\rangle _{\text{2D}}=-\alpha \omega _{\text{LO}}%
{\displaystyle{\hbar \omega _{\text{LO}} \over m_{b}}}%
\sqrt{%
{\displaystyle{\hbar  \over m_{b}\omega _{\text{LO}}}}%
}%
{\displaystyle{1 \over \sqrt{2}}}%
\displaystyle\int %
\limits_{0}^{\infty }dk%
\displaystyle\int %
\limits_{0}^{\infty }d\omega \text{ }k^{2}%
{\displaystyle{S_{\text{2D}}({\bf k},\omega -\omega _{\text{LO}}) \over \omega ^{2}}}%
.  \eqnum{19b}  \label{w2d}
\end{equation}
The 2D and 3D results are related by the scaling relation 
\begin{equation}
\left\langle \omega \right\rangle _{\text{2D}}(\alpha ,S_{\text{2D}%
})=\left\langle \omega \right\rangle _{\text{3D}}(3\pi \alpha /4,S_{\text{3D}%
})  \label{scaling}
\end{equation}
provided the corresponding 2D or 3D dynamical structure factor of the
electron (hole) system is used.\ Taking the low-density limit of (\ref
{scaling}), the correct one-polaron scaling relation \cite{WuPRB36} is
retrieved. The present analysis shows that the generalization (\ref{scaling}%
) of the one-polaron scaling relation holds in the many-polaron case,
provided that the corresponding 2D or 3D dynamical structure factor is used.
Both expressions (\ref{w3d}) and (\ref{w2d}) contain the Fr\"{o}hlich
electron-phonon (or hole-phonon) coupling constant $\alpha $ only as a
prefactor: the remaining integrations of (\ref{w3d}) and (\ref{w2d}) depend
only on the details of the electron-electron (or hole-hole) many-body
effects.

\subsection{Results and discussion}

For one polaron, the normalized first frequency moments are {$\left\langle
\omega \right\rangle _{\text{3D}}=(2/3)\left\langle \omega \right\rangle _{%
\text{2D}}$} and $\left\langle \omega \right\rangle _{\text{2D}}=(\pi
/2)\alpha \omega _{\text{LO}}$ \cite{DevreeseEAP14}. These limits correspond
to the low density limit of the expressions (\ref{w3d}) and (\ref{w2d}),
which is taken by letting $k_{\text{F}}\rightarrow 0$. In this limit the
dynamical structure factor $S({\bf k},w)$ becomes strongly peaked along $%
w=\hbar k^{2}/(2m_{b})$. In the case of many polarons, many-body effects
such as the electron-electron interaction, screening effects, and the
occupational effect due to Fermi statistics will play a role.

Figure 1 shows the result obtained from (\ref{w3d}),(\ref{w2d}) for the
first frequency moment of the optical absorption in 3D and 2D as a function
of density, expressed through the Fermi wave vector $k_{\text{F}}$. For the
3D case, the material parameters of ZnO were used ($\omega _{\text{LO}}=73.27
$ meV, $\varepsilon _{\infty }=4.0$, $m_{b}=0.24$ $m_{e}$) and for the 2D
case, the material parameters for GaAs were taken ($\omega _{\text{LO}}=36.77
$ meV, $\varepsilon _{\infty }=10.9$, $m_{b}=0.0657$ $m_{e}$) \cite
{Greenbook}. The limit of low density corresponds to the one polaron-results
indicated by arrows starting on the $y$-axis.

The dashed curve (for 2D) and the dotted curve (for 3D) were calculated with
the Hartree-Fock structure factor of the electron (hole) system. The
increase of $\left\langle \omega \right\rangle $ in the region $k_{\text{F}%
}\lessapprox 1$ $\sqrt{m_{b}\omega _{\text{LO}}/\hbar }$ shows an initial
shift in spectral density towards higher frequencies. The full curve (for
2D) and the dash-dotted curve (for 3D) are the result obtained by
substituting the RPA structure factor of the electron (hole) system in
expressions (\ref{w3d}),(\ref{w2d}). When the many-body effects are taken
into account on the level of the RPA, they result in a monotonously
decreasing value of the normalized first frequency moment as a function of
density.

The results obtained from the sum rule expressions (\ref{w3d}),(\ref{w2d})
are in agreement with the result \cite{TemperePRBs} obtained by direct
integration of the many-polaron optical absorption spectrum calculated with
the variational ground state wave function of Lemmens, Brosens and Devreese 
\cite{LemmensPSS82}. Note that the derivation presented in the present paper
does not rely on the variational wave function used by LDB. The use of the
force-force correlation function introduces already a factor $\alpha $ in
expression (\ref{<w>bis}), so that it is sufficient to calculate the
expectation value in (\ref{<w>bis}) with respect to the phonon vacuum in
order to obtain the first frequency moment up to order $\alpha $ in the
charge carrier - phonon coupling strength.

\section{Zeroth frequency moment and many-body effects in the polaron mass}

In the polaron optical absorption spectra at zero temperature, a
delta-function contribution is present at $\omega =0$ \cite{DevreesePRB15}.
This feature of the polaron spectrum will, at low temperatures, elude
experimental detection since all infrared absorption experiments have
naturally a lowest measurable frequency. Nevertheless, this delta-function
contribution will play a role in the zeroth frequency moment of the optical
absorption and hence in all normalized frequency moments. Note that as the
temperature is raised, this DC delta-function contribution will
substantially widen and become detectable \cite{HuybrechtsSSC13}. At
temperature zero, we must rely on an adaptation of the f-sum rule for one
polaron, derived in \cite{DevreesePRB15}: 
\begin{equation}
{\displaystyle{N\pi e^{2} \over 2m^{\ast }}}%
+%
\displaystyle\int %
\limits_{\omega _{\text{LO}}}^{\infty }%
\mathop{\rm Re}%
[\sigma (\omega )]d\omega =%
{\displaystyle{N\pi e^{2} \over 2m_{b}}}%
.  \label{DLR}
\end{equation}
In this expression, $m^{\ast }$ is the polaron effective mass. At very low
temperature, the integral in the left-hand-side of (\ref{DLR}) can be
determined experimentally or used to derive the effective polaron mass from
many-polaron optical absorption theories such as refs. \cite
{TemperePRBs,CataudellaEJPB12}. The LO phonon frequency $\omega _{\text{LO}}$
is typically a few hundred cm$^{-1}$, obviously within the range of far
infrared spectroscopy. Expression (\ref{DLR}) allows then to determine the
polaron mass from the optical conductivity: 
\begin{equation}
1-%
{\displaystyle{m_{b} \over m^{\ast }}}%
=%
{\displaystyle{2m_{b} \over N\pi e^{2}}}%
\displaystyle\int %
\limits_{\omega _{\text{LO}}}^{\infty }%
\mathop{\rm Re}%
[\sigma (\omega )]d\omega .  \label{SR0}
\end{equation}
The evolution of the polaron effective mass as a function of the density of
the interacting many-polaron gas is shown in Figure 2. To obtain the polaron
effective mass, the sum rule (\ref{SR0}) was applied to the many-polaron
optical absorption spectrum calculated in \cite{TemperePRBs}: 
\begin{equation}
\mathop{\rm Re}%
[\sigma (\omega )]=\frac{n}{\hbar \omega ^{3}}\frac{e^{2}}{m_{b}^{2}}\sum_{%
{\bf k}}k_{x}^{2}|V_{{\bf k}}|^{2}S({\bf k},\omega -\omega _{\text{LO}})
\end{equation}
The result for $(m^{\ast }/m_{b}-1)/\alpha $ is shown in Figure 2 as a
function of density expressed via $k_{\text{F}}$. For low density the
effective mass naturally converges to its one-polaron value in the weak and
intermediate electron-phonon coupling limit, given by $m^{\ast
}=m_{b}(1+\alpha /6)$ in 3D and by $m^{\ast }=m_{b}(1+\pi \alpha /8)$ in 2D 
\cite{DevreeseEAP14}. Both dashed curves were calculated using the
Hartree-Fock structure factor. The shift of spectral weight towards higher
frequencies, due to occupation effects, is reflected here as an initial
increase in the effective polaron mass. The full and dash-dotted curves show
the result using the RPA structure factor. The resulting effective polaron
mass is a monotonously decreasing function of density.

The normalized first frequency moment of the optical absorption, {\it %
excluding }the delta-function contribution, $\left\langle \omega
\right\rangle _{0}$, is accessible to experiment, using: 
\begin{equation}
\left\langle \omega \right\rangle _{0}=%
{\displaystyle{%
\displaystyle\int \limits_{\omega _{\text{LO}}}^{\infty }\omega %
\mathop{\rm Re}[\sigma (\omega )]d\omega  \over %
\displaystyle\int \limits_{\omega _{\text{LO}}}^{\infty }%
\mathop{\rm Re}[\sigma (\omega )]d\omega }}%
.  \label{w0}
\end{equation}
The question of the high-frequency cutoff in the experiment will be
discussed as a further refinement in the next section. The numerator in
expression (\ref{w0}) is still coincident with the numerator in (\ref{<w>})
for many-polaron optical conductivity. Based on (\ref{SR0}) for the
denominator, we obtain 
\begin{equation}
\left\langle \omega \right\rangle =\left( 1-%
{\displaystyle{m_{b} \over m^{\ast }}}%
\right) \left\langle \omega \right\rangle _{0}.
\end{equation}
As a consequence, to obtain the normalized first frequency moment $%
\left\langle \omega \right\rangle $ of the optical absorption including the
delta function contribution, one has to correct the result $\left\langle
\omega \right\rangle _{0}$ obtained without the delta function, by
multiplying it with a correction factor $\left( 1-%
{\displaystyle{m_{b} \over m^{\ast }}}%
\right) $.

\section{High frequency tail of the many-polaron optical absorption spectrum}

An upper bound $\omega _{\text{max}}$ of the integration domain (cutoff
frequency) is necessarily present in the experimental determination of the
frequency integrals in (\ref{w0}). Thus the experimentally accessible
quantity for the experiment is 
\begin{equation}
\left\langle \omega \right\rangle _{\text{exp}}=%
{\displaystyle{%
\displaystyle\int \limits_{\omega _{\text{LO}}}^{\omega _{\text{max}}}\omega %
\mathop{\rm Re}[\sigma (\omega )]d\omega  \over %
\displaystyle\int \limits_{\omega _{\text{LO}}}^{\omega _{\text{max}}}%
\mathop{\rm Re}[\sigma (\omega )]d\omega }}%
,
\end{equation}
where $\omega _{\text{max}}$ is typically of the order of 10000 cm$^{-1}$ 
\cite{LupiPRL83}. This upper bound will influence the experimental value of
the first moment $\left\langle \omega \right\rangle $ since a high-frequency
tail ($\omega >\omega _{\text{max}}$) is present in the optical absorption
spectrum of polarons.

Figure 3 illustrates this cutoff problem. The full curve in Fig. 3 shows the
sum rule result for the normalized first frequency moment of the optical
absorption of the many-polaron system as a function of the Fermi wave
vector. The other curves show the results, obtained by direct integration of
the spectrum calculated with the variational ground state wave function of
ref. \cite{TemperePRBs}, using different values for the cutoff parameter $%
\omega _{\text{max}}$. Already at $\omega _{\text{max}}=10$ $\omega _{\text{%
LO}}$ a considerable difference is found between the result with cutoff and
the result without cutoff.

For energy transfers $\hbar \omega $ much larger than the Hartree-Fock
exchange energy of the many-polaron gas, the optical absorption will not be
strongly modified by many-body effects. This property is used in the optical
absorption measurements to extrapolate the high-frequency measurements of
the reflectance by using a free-electron response (as described e.g. in ref. 
\cite{ZhangPRB43}). For the many-polaron optical absorption in the high
frequency limit (for example in 3D) we find \cite{TemperePRBs}:\ 
\begin{equation}
\lim\limits_{\omega \rightarrow \infty }%
\mathop{\rm Re}%
[\sigma (\omega )]=%
{\displaystyle{Ne^{2} \over m_{b}}}%
\text{ }\frac{2}{3}\alpha 
{\displaystyle{\sqrt{\omega -1} \over \omega ^{3}}}%
,
\end{equation}
so that 
\begin{equation}
\lim\limits_{\omega _{\text{max}}\rightarrow \infty }%
\displaystyle\int %
\limits_{\omega _{\text{max}}}^{\infty }\omega 
\mathop{\rm Re}%
[\sigma (\omega )]=%
{\displaystyle{Ne^{2} \over m_{b}}}%
\frac{2}{3}\alpha \left[ 
{\displaystyle{\sqrt{\omega _{\text{max}}-1} \over \omega _{\text{max}}}}%
+\arcsin \left( 
{\displaystyle{1 \over \sqrt{\omega _{\text{max}}}}}%
\right) \right] .
\end{equation}
For the first frequency moment this leads to 
\begin{equation}
\left\langle \omega \right\rangle =\lim\limits_{\omega _{\text{max}%
}\rightarrow \infty }\left\{ \left\langle \omega \right\rangle _{\text{exp}%
}(\omega _{\text{max}})+%
{\displaystyle{4 \over 3\pi }}%
\alpha \left[ 
{\displaystyle{\sqrt{\omega _{\text{max}}-1} \over \omega _{\text{max}}}}%
+\arcsin \left( 
{\displaystyle{1 \over \sqrt{\omega _{\text{max}}}}}%
\right) \right] \right\} .  \label{cutoff}
\end{equation}
Expression (\ref{cutoff}) allows one to correct the error made by using a
cutoff frequency in determining the experimental value for the first
frequency moment of the polaron optical absorption. How large does $\omega _{%
\text{max}}$ have to be for the limit (\ref{cutoff}) to be accurate ?
Naturally, this depends on the density. The dynamical structure factor of
the electron (hole) system reduces to the dynamical structure factor of a
one-particle system for $k\gg 2k_{\text{F}}$, as does the static structure
factor \cite{Gorobchenko}. At these values of the wave vector, the only
frequency for which the dynamical structure factor $S(k,\omega -\omega _{%
\text{LO}})$ differs substantially from zero is for $\omega -\omega _{\text{%
LO}}=k^{2}/2$. Hence, we can estimate that the many-polaron optical
absorption will be well approximated by the one-polaron optical absorption
only for frequencies $\omega \gg 2k_{F}^{2}+\omega _{\text{LO}}$. This is
also the value of $\omega _{\text{max}}$ which should be used as
high-frequency cutoff in the experiments determining $\left\langle \omega
\right\rangle _{\text{exp}}$. For example, in ZnO ($\omega _{\text{LO}}=73.27
$ meV, $\varepsilon _{\infty }=4.0$, $m_{b}=0.24$ $m_{e}$) with a charge
carrier density of $n=10^{20}$ cm$^{-3}$, this would correspond to $\omega _{%
\text{max}}\approx 18.9$ $\omega _{\text{LO}}$ or $\omega _{\text{max}%
}\approx 11000$ cm$^{-1}$.

\section{Conclusion}

Expressing the optical conductivity as a force-force correlation function, a
closed expression is obtained for the sum rule of the normalized first
moment of a many-polaron system, in such a way that the electron-phonon
coupling and the many-body effects of the electron (hole) system formally
decouple. This procedure allows one to treat the many-body effects in the
system of charge carriers using any desired approximation (Hartree-Fock,
RPA, ...). The results obtained by the sum rule derived here are valid for a
many-polaron system with weak electron-phonon coupling and at temperature
zero.

Experimental data are characterized by a cutoff frequency both for low and
high frequencies This in general complicates the use of any sum rule, since
its evaluation in principle requires a summation over all frequencies, $%
\omega =0...\infty $. To overcome this problem we introduced formula (\ref
{cutoff}) for many-polaron optical absorption spectra at high frequencies,
and discussed up to which frequency experimental data have to be available
so that this extrapolation formula is useful. At zero temperature, a delta
function contribution is present at $\omega =0$ in the many-polaron optical
absorption spectrum \cite{DevreesePRB15}. The spectral weight of this delta
function contribution is related to the polaron mass \cite{DevreesePRB15},
which in its turn can be derived from a measurement of the zeroth frequency
moment of the many-polaron optical absorption.

\section{Acknowledgments}

The authors like to acknowledge S. N. Klimin and V. M. Fomin for helpful
discussions and intensive interactions. We thank P. Calvani for fruitful
discussions. One of us, J.T., (``Postdoctoraal Onderzoeker van het Fonds
voor Wetenschappelijk Onderzoek -- Vlaanderen''), is supported financially
by the Fonds voor Wetenschappelijk Onderzoek -- Vlaanderen (Fund for
Scientific Research -- Flanders). Part of this work is performed in the
framework of the ``Interuniversity Poles of Attraction Program -- Belgian
State, Prime Minister's Office -- Federal Office for Scientific, Technical
and Cultural Affairs'' (``Interuniversitaire Attractiepolen -- Belgische
Staat, Diensten van de Eerste Minister -- Wetenschappelijke, Technische en
Culturele Aangelegenheden''), and in the framework of the FWO projects
1.5.545.98, G.0287.95, 9.0193.97, WO.025.99N and WO.073.94N
(Wetenschappelijke Onderzoeksgemeenschap, Scientific Research Community of
the FWO on ``Low Dimensional Systems''), and in the framework of the BOF NOI
1997 and GOA\ BOF UA 2000 projects of the Universiteit Antwerpen.

\bigskip

\section*{Figure captions}

{\bf Figure 1}: The normalized first frequency moment of the optical
absorption of a gas of polarons including many-body effects in the system of
constituent charge carriers, as obtained with the present treatment, is
shown as a function of density, expressed through the Fermi wave vector $k_{%
\text{F}}$ in polaron units. The top two curves are for the 2D polaron gas
(with parameters corresponding to GaAs: $\omega _{\text{LO}}=36.77$ meV, $%
\varepsilon _{\infty }=10.9$, $m_{b}=0.0657$ $m_{e}$), and the bottom two
curves are the results for the 3D case (with parameters representative of
ZnO: $\omega _{\text{LO}}=73.27$ meV, $\varepsilon _{\infty }=4.0$, $%
m_{b}=0.24$ $m_{e}$).

{\bf Figure 2}: The effective polaron mass in the many-polaron system is
shown as a function of density, for the 2D case (GaAs parameters: $\omega _{%
\text{LO}}=36.77$ meV, $\varepsilon _{\infty }=10.9$, $m_{b}=0.0657$ $m_{e}$%
) and the 3D case (ZnO parameters: $\omega _{\text{LO}}=73.27$ meV, $%
\varepsilon _{\infty }=4.0$, $m_{b}=0.24$ $m_{e}$), and for different
approximations (Hartree-Fock and RPA) to take the many-body effects into
account.

{\bf Figure 3}: The effect of introducing an upper limit (a cutoff, $\omega
_{\text{max}}$) to the integration domain in the expression for the first
frequency moment $\left\langle \omega \right\rangle $ of the optical
absorption of the polaron gas is shown in this figure. The different curves
show $\left\langle \omega \right\rangle /\left\langle \omega \right\rangle
_{k_{\text{F}}=0}$ as a function of the Fermi wave vector for the set of
cutoff values listed in the legend, obtained from direct integration of the
many-polaron spectrum as calculated in \cite{TemperePRBs}. The full curve
shows the sum rule result (without cutoff) in the present treatment. In the
inset, the absolute value of $\left\langle \omega \right\rangle $ is shown
as a function of the Fermi wave vector, for the same set of cutoff values.

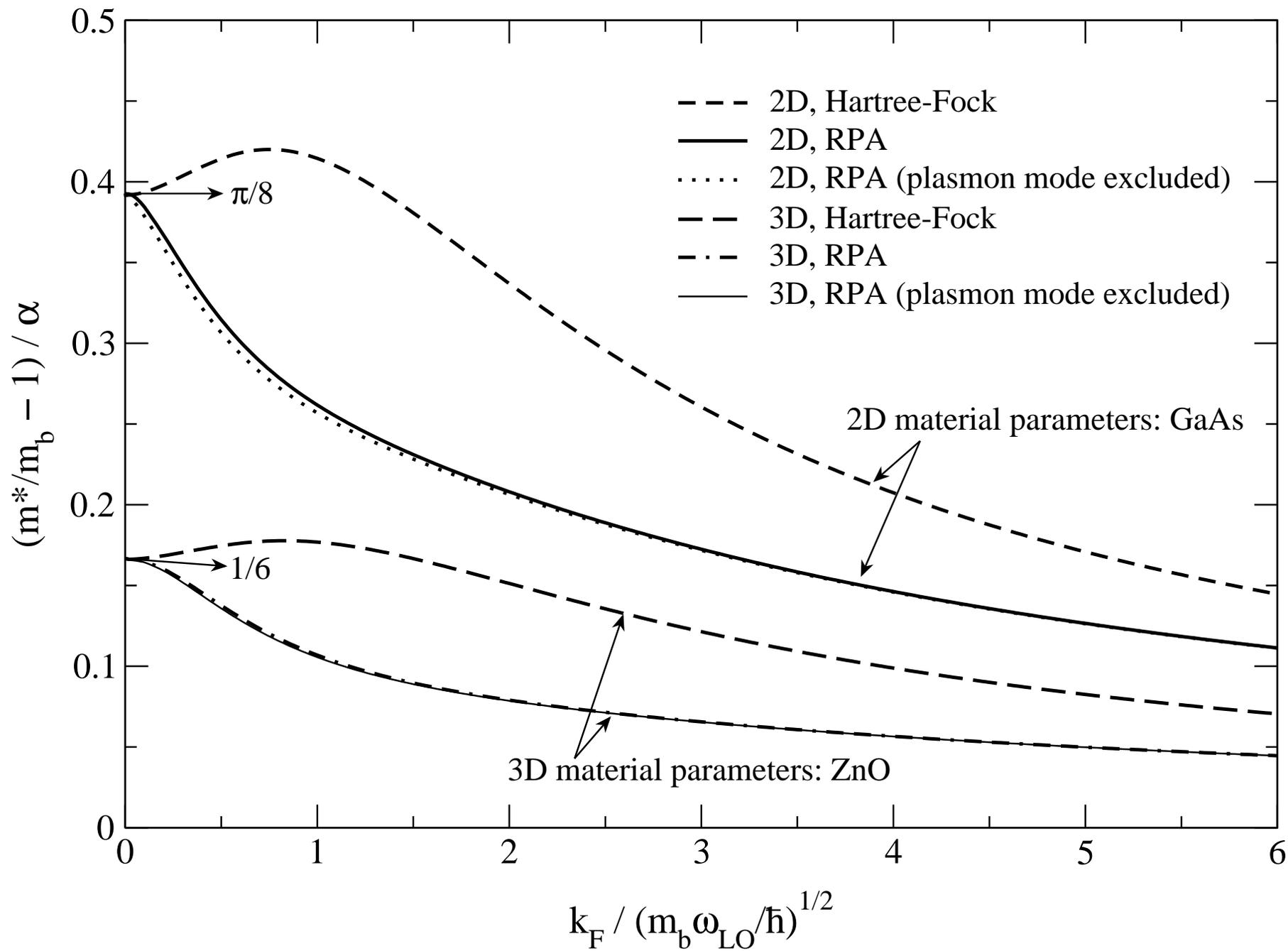
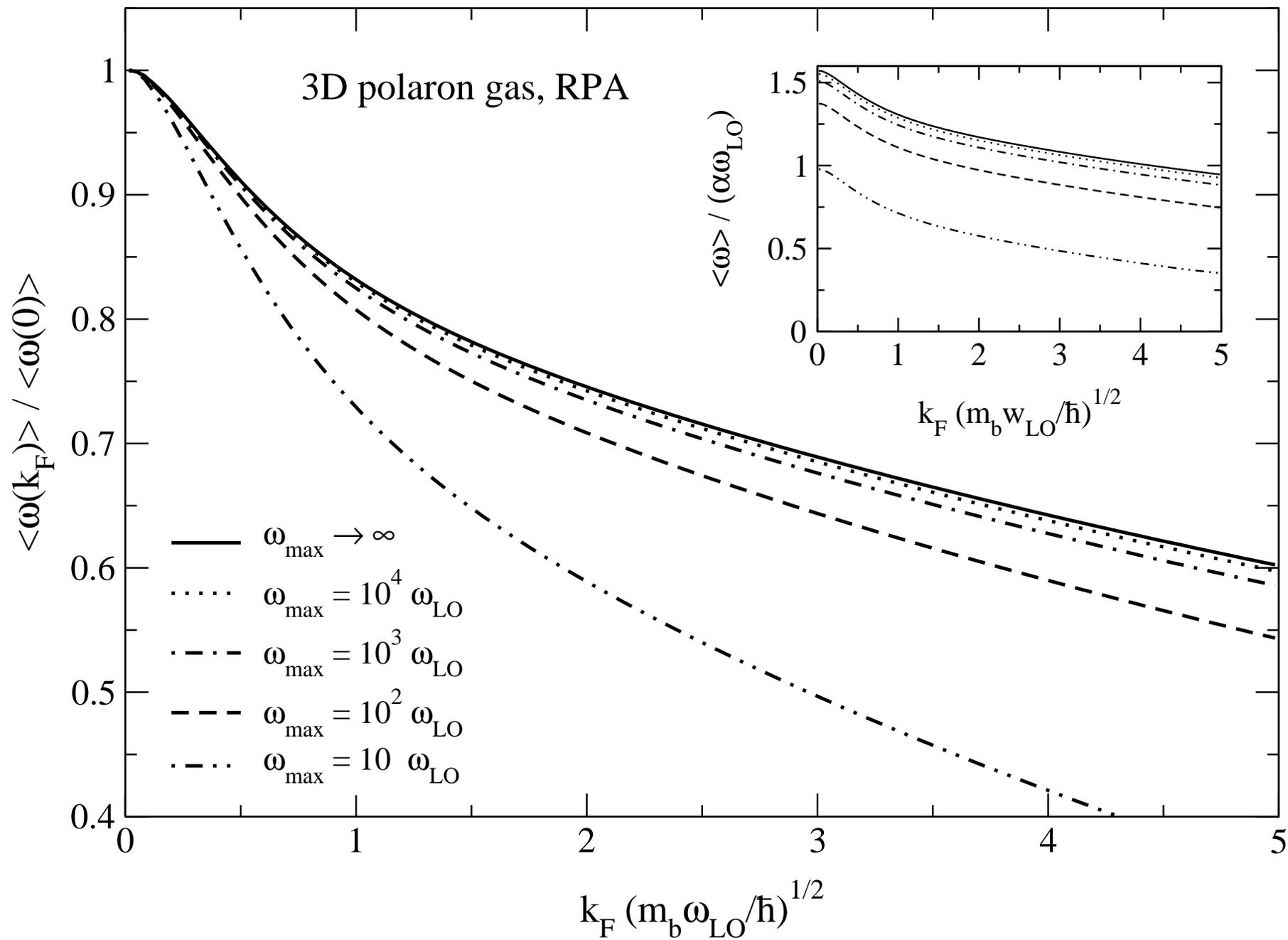
\end{document}